# Sorting Reordered Packets with Interrupt Coalescing

Wenji Wu, Phil Demar, Matt Crawford
wenji@fnal.gov, demar@fnal.gov, crawdad@fnal.gov
Fermilab MS-120, Batavia, IL 60510 USA,
Phone: +1 630-840-4541, Fax: +1 630-840-8208

*Abstract* -- TCP performs poorly in networks with serious packet reordering. Processing reordered packets in the TCP layer is costly and inefficient, involving interaction of the sender and receiver. Motivated by the interrupt coalescing mechanism that delivers packets upward for protocol processing in blocks, we propose a new strategy, Sorting Reordered Packets with Interrupt Coalescing (SRPIC), to reduce packet reordering in the receiver. SRPIC works in the network device driver; it makes use of the interrupt coalescing mechanism to sort the reordered packets belonging to the same TCP stream in a block of packets before delivering them upward; each sorted block is internally ordered. Experiments have proven the effectiveness of SRPIC against forward-path reordering.

## 1. Introduction and Motivation

The last 30 years have witnessed the tremendous success of the Internet Protocol suite. It was developed by the Defense Advanced Research Projects Agency (DARPA), and has been used widely in military, educational, and commercial systems. The fundamental architectural feature of the Internet is the use of datagrams (packets) as the units which are transported across the underlying networks individually and independently; the datagram provides a basic building block on which a variety of types of services can be implemented [1]. The widely deployment of TCP/IP has been attributed to this feature. However, datagrams can arrive at the destination out-of-sequence, necessitating packet reordering, and degrading the performance of high-layer services such as TCP.

Internet measurement studies [2][3][4][5][6][7][8][9][10] have shown that the phenomenon of packet reordering exists throughout the Internet and sometimes can be severe. Causes of packet reordering in IP networks have been identified in [10][11][12], including packet-level multi-path routing, route flapping, inherent parallelism in high-speed routers, link-layer retransmission, and router forwarding lulls. Packet reordering is now one of the four metrics describing QoS in packet networks, along with delay, loss, and jitter.

Two trends regarding packet reordering need to be emphasized. First, the studies in [3][9][10] demonstrate a strong correlation between inter-packet spacing and packet reordering. With the deployment of high-speed TCP variants, such as FAST TCP[13], CUBIC[14], BIC[15], and HSTCP[16], sustained high TCP throughput has become achievable in very high bandwidth networks. The smaller inter-packet spacing resulting from high throughput may increase the probability of packet reordering. Second, local parallelism is on the increase within the Internet because it reduces equipment and trunk costs [4]. Backbone network link technology has reached 10 Gbps, with 40 Gbps or 100 Gbps on the horizon. More and more parallelism is being introduced into the network equipments to reduce cost or minimize engineering difficulties. Even the largest network equipment vendors cannot avoid troubles with reordering in their network devices [10].

For connection-oriented reliable data transmission, packet reordering is dealt with in the TCP layer of the Internet architecture [1]. TCP performs poorly in networks with severe packet reordering. Studies in [4][11] clearly discuss the impact of packet reordering on TCP. Over the years, various reordering-tolerant algorithms [17][18][19][20][21][22] have been proposed to deal with packet reordering in TCP. Studies in [11] [23] have proven the effectiveness of these reordering-tolerant algorithms. One common characteristic of these algorithms is that they all operate in the TCP layer and react passively to out-of-sequence packets, instead of behaving actively to eliminate or reduce packet reordering to save processing in TCP, which is costly and involves interaction of sender and receiver. So far, little research on actively eliminating or reducing packet reordering at lower network layers has been reported and published. Motivated by the interrupt coalescing mechanism [24], which delivers packets in blocks for higher layer protocol processing, we propose a new strategy to eliminate or reduce the packet reordering in the receiver, which we call Sorting Reordered Packets with Interrupt Coalescing (SRPIC). SRPIC works in the network device driver. It makes use of the interrupt coalescing mechanism to sort reordered packets of the same TCP stream in a block of packets before delivering them upward; each sorted block is internally ordered. Experiments have demonstrated the effectiveness of SRPIC against forward-path reordering. The benefits of SRPIC are: (1) saving costly processing in the TCP layer in both the sender and the receiver, increasing overall system efficiency; (2) achieving higher TCP throughput; (3) maintaining TCP self-clocking while avoiding injection of bursty traffic into the network; (4) reducing or eliminating unnecessary retransmissions and duplicate ACKs or SACKs in the network, enhancing overall network efficiency; and (5) coexistence with other packet reordering-tolerant algorithms. However, it should be emphasize that SRPIC is a mechanism in the device driver that complements the TCP layer, instead of replacing it; packet reordering that can not be eliminated by SRPIC will be finally processed by TCP.

The remainder of the paper is organized as follows: In Section 2, background and related research on TCP packet reordering is presented. Section 3 describes the SRPIC algorithm. In section 4, we present experiment results on the effectiveness of SRPIC. And finally in section 5, we conclude the paper.

## 2. Background & Related Works

### 2.1 Interrupt Coalescing

Most operating systems deployed on the network, such as FreeBSD, Solaris, Linux, and Windows [25][26][27][28], are interrupt-driven. When packets are received, the network interface card (NIC) typically interrupts to inform the CPU that packets have arrived. Without some form of interrupt moderation logic on the network device driver, this might lead to an interrupt for each incoming packet. As the traffic rate increases, the interrupt operations become very costly.

Interrupt coalescing was first proposed by J.C. Mogul et al. in [24]. The idea of interrupt coalescing is to avoid flooding the host system with too many NIC interrupts. Each interrupt serviced may result in the processing of several received packets. The system gets more efficient as the traffic load increases. Usually the interrupt coalescing mechanism works as follows [24][29]: incoming packets are first transferred into the ring buffer, and then the NIC raises a hardware interrupt. When CPU responds to the interrupt, the corresponding interrupt handler is called, within which a deferred procedure call (DPC) (Windows) [28], or a softirq (Linux) [27], is scheduled. At the same time, the NIC's receive interrupt function is disabled. DPC or softirq is serviced shortly after and moves packets from ring buffer upward for higher layer protocol processing till the ring buffer is empty. After that, DPC or softirq enables the NIC interrupt and exits. When more packets come, the cycle repeats.

### 2.2 Impact of Packet Reordering on TCP

TCP is a reliable transport protocol, designed to recover from misbehavior at the Internet Protocol (IP) layer. The details of TCP protocol are specified in [30]. TCP performs poorly in networks with severe packet reordering. Studies in [4] [11] discuss in detail the impact of packet reordering on TCP. The impact of packet reordering on TCP is multifold:

- Degrading the receiver efficiency: many TCP implementations use the header prediction algorithm [31] to reduce the costs of TCP processing. However, header prediction only works for in-sequence TCP segments. If segments are reordered, most TCP implementations do far more processing than they would for in-sequence delivery. The two cases are usually termed fast path, and slow path respectively. When TCP receives packets in sequence, it will stay on the fast path, and simply acknowledge the data as it's received. Fast path processing has sequential code with well-behaved branches and loops; CPU cache can have nearly perfect efficiency. However, if the received segments are out of order, the receiving TCP will be processing in the slow path and duplicate acknowledgements will be generated and sent; if the TCP has selective acknowledgements (SACK) [32] and duplicate SACK (DSACK) [18] enabled, the receiver will sort the out-of-order queue to generate SACK blocks. Sorting the out-of-order queue is expensive, especially when the queue is large. Slow path processing leads to a random pattern of data access, which is far less deterministic and presents a challenge for CPU caches. Packet reordering places serious burdens on the TCP receiver.

- Degrading the sender efficiency: when duplicate acknowledgements (dupACKs) come back to the sender, the sender TCP will also be processing in the slow path. If dupACKs include SACK options, the computational overhead of the processing SACK block is high. Packets in flight and not yet acknowledged are held in the retransmission queue. On receipt of SACK information, the retransmission queue would be walked and the relevant packets tagged as sacked or lost. For large bandwidth-delay products [29], the retransmission queue is very large, and the walk is costly. Finally, if the number of dupACKs exceeds *dupthresh,* the sending TCP may go to fast retransmit and perform unnecessary retransmissions. The TCP reordering-tolerant algorithms may also adjust *dupthresh* to avoid unnecessary fast retransmit or rapid recovery from the false reductions of *cwnd* and *ssthresh*. These operations are also costly.

- Degrading TCP performance: most TCP implementations consider three or more duplicate ACKs as an indication that a packet has been lost, based on the assumption that reordered packet can trigger only one or two duplicate ACKs. However, if packets are reordered to a slightly greater degree, TCP misinterprets it as a lost packet and unnecessarily invokes fast retransmit/fast recovery to reduce the congestion window in the sender. Thus the congestion window may be kept small relative to the available bandwidth of the path with persistent and substantial packet reordering. Packet reordering would lead to loss of TCP self-clocking and understating of estimated RTT and RTO, which also throttle the TCP throughput.
- Wasting Network bandwidth: The unnecessary retransmissions in the forward path and the dupACKs/SACKs in the reverse path waste the network bandwidth.

**2.3 Related Works**

Over the years, various reordering-tolerant algorithms have been proposed for packet reordering in TCP. The most widely deployed algorithms are the Eifel algorithm [17] by Ludwig and Katz, and the DSACK TCP [18] by S. Floyd. K. Leung et al. give a comprehensive survey of reordering-tolerant algorithms in [11]. In general, those algorithms either adaptively adjust TCP reordering threshold *dupthresh* to avoid false fast retransmit, or rapidly recover from the false reductions of congestion window *cwnd* and slow start threshold *ssthresh* in the sender. There is also a different group of reordering-tolerant algorithms termed *Response Postponement* in [11]. The response postponement algorithms avoid triggering spurious congestion responses in the sender or duplicate ACKs generation in the receiver by deferring them for a time period, in the hope that the out-of-sequence packets might come during the period. From this perspective, the response postponement algorithms are similar to our proposed SRPIC. One common characteristic of these algorithms is that they operate in the TCP layer and react passively to out of sequence packets, instead of behaving actively to eliminate or reduce packet reordering to save processing in TCP.

SRPIC is different from these algorithms in a number of ways. First, it operates in the low layer of the protocol stack, and avoids complicating the already bloated TCP layer. In addition, the SRPIC mechanism could be applied to other services, not just TCP. Secondly, SRPIC behaves actively to eliminate or reduce packet reordering to save processing in TCP. Finally, since interrupt coalescing can naturally group packets into blocks, SRPIC does not require timing mechanisms. By contrast, response postponement algorithms require timing mechanisms for the deferred operations, complicating TCP implementation.

In [33], S. Govind et al. proposed a packet sort mechanism to reduce packet reordering in network processors. Their work can be applied in network devices such as routers and switches. So far, no research has been found to actively eliminate or reduce packet reordering in the receiver to save the costly packet reordering processing in the TCP layer.

### 3. Sorting Reordered Packets with Interrupt Coalescing (SRPIC)

#### 3.1 Interrupt coalescing packet block size

The interrupt coalescing mechanism handles multiple packets per interrupt as the packet rate increases. In each interrupt, the packets belonging to the same stream make up a block. In this section, we study the relationship of the block size to the incoming packet rate. In general, the packet receiving process of different OSes are similar, but use different terminologies. For example, the softirq in Linux is called Deferred Procedure Call in Windows. In the following analysis, we assume the receiver to be running Linux.

Assume there is bulk data flowing from sender to receiver, such as an FTP file transfer. Process A is the data receiving process in the receiver. For simplicity, assume only process A runs on the receiver, and no other traffic is being directed to it. At time $t$, packets are incoming at the receiver with a rate of $P(t)$ packets/second (pps). Also, let $T_{intr}$ be the interrupt-coalescing driver's hardware interrupt delay, which includes NIC interrupt dispatch and service time; the software interrupt *softnet*'s packet service rate be $R_{sn}$ (pps).

At time $t_0 - \varepsilon$, the ring buffer is empty and A is waiting for network data from the sender. At time $t_0$, packets start to arrive in the receiver. As an interrupt-driven operating system, the OS execution sequence is: hardware interrupt → software interrupt (or DPC) → process [27][34]. Arriving packets are first transferred to ring buffer. The NIC raises a hardware interrupt which results in scheduling the softirq softnet. The softnet handler starts to move packets from ring buffer to the socket receive buffer of process A, waking up process A and putting it into the run queue. During this period, new packets might arrive at the receiver. Softnet continues to process the ring buffer till it is empty. Then softirq yields the CPU. Process A begins to run, moving data from the socket's receive buffer into user space. Typically, process A runs out of data before the next packet arrives at the receiver, and goes to sleep, waiting for more. If the next packet always arrives before process A goes to sleep, the sender will overrun the receiver. Incoming packets would accumulate in the socket's receive buffer. For TCP traffic, the flow control mechanism would eventually take effect to slow down the sender. When the next packet arrives at the receiver, the sequence of events is repeated. The cycle repeats until process A stops. Figure 1 illustrates the packet receiving process with interrupt coalescing.

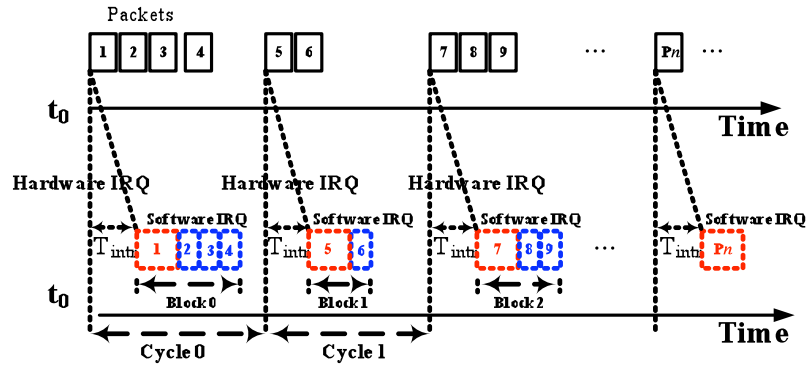

**Figure 1 Packet Receiving Process with Interrupt Coalescing**

We use $j$ to denote cycles; cycle $j$ starts at time $t_j$. Letting $T_{sn}^j$ be the time that softnet spends emptying the ring buffer in cycle $j$, we see that

$$1 + \left\lfloor \int_{t_j}^{t_j+T_{intr}+T_{sn}^j} P(t)d(t) \right\rfloor = T_{sn}^j * R_{sn} \tag{1}$$

Here, $T_{sn}^j * R_{sn}$ is actually the number of packets that are handled together in cycle $j$. We call this group of packets as block $j$; the block size $Block^j$ is:

$$Block^j = T_{sn}^j * R_{sn} \tag{2}$$

For any given receiver, $T_{intr}$ and $R_{sn}$ are relatively fixed. Based on (1) and (2), it is clear that the block size will increase nonlinearly with the data rate $P(t)$. For example, if $P(t)$ is relatively stable in cycle $j$, with average value $P^j$, then we will have

$$Block^j = \left\lfloor \frac{1 + T_{intr} * P^j}{R_{sn} - P^j} * R_{sn} \right\rfloor \tag{3}$$

To illustrate the relationship between the block size and $P(t)$, we run data transmission experiments over an isolated sub-network. In the experiments, we run *iperf* [35] to send data in one direction between two computer systems. The sender and receiver's detailed features are as shown in Section 4. The Round Trip Time (RTT) statistics are: *min/avg/max/dev = 0.134/0.146/0.221/0.25 ms,* with no packet loss. We use three different NICs (100Mbps, 1Gbps, and 10Gbps respectively) in the sender to vary the bandwidth and control $P(t)$. Iperf runs for 50 seconds; we record the throughput and the block size at different cycles. Figure 2 illustrates the block sizes with different throughput[*]. Figure 2.A, 2.B, and 2.C correspond to the sender NIC 100Mbps, 1Gbps, and 10Gbps respectively. Figure 2 clearly shows that the block size increases with the throughput: when the throughput is low at 95.3Mbps, each block has only one packet (Figure 2.A); at a throughput rate of around 938Mbps, the block size is commonly around 7 packets (Figure 2.B); when the throughput is high at 5.89 Gbps, blocks typically have 30 to 60 packets (Figure 2.C).

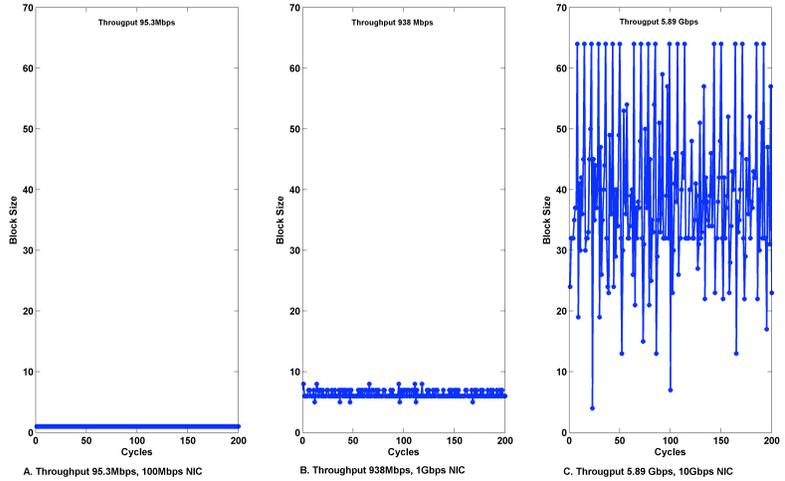

**Figure 2 Block Sizes vs. Different Incoming Data Rates**

If multiple streams are transmitting to the same receiver, the "effective" $R_{sn}$ for each stream is actually decreasing. But for each individual stream, the relationship between the block size and $P(t)$ remains the same: the block size will increase as the data rate $P(t)$ is raised.

## 3.2 Sorting Reordered Packets with Interrupt Coalescing (SRPIC)

Following the TCP/IP design philosophy [1], most network device drivers deliver received packets upward for higher layer processing in the same sequence as they are received. The lower layers of the protocol (below the transport layer) do not take any measures to process packet reordering, other than reassembling IP fragments. Packet reordering is dealt with either in the TCP layer for connection-oriented reliable data transmission, or in the application layer for other transports. For example, some multimedia applications like VOIP use a jitter buffer in the application layer to reorder and smooth out packets. In this paper, we focus on TCP packet reordering. As discussed above, reordered packets lead to extra TCP processing overheads in both the sender and the receiver: slow-path processing, DupACKs/SACK generation, and the computational overhead in processing SACK blocks. In the worst case, packet reordering will lead to false fast retransmit, resulting in small congestion window in the sender and severely degrading throughput.

---

[*] For the illustration's purpose, only 200 consecutive cycles' data are shown.

Motivated by the fact that with interrupt coalescing, the network device driver delivers the received packets in blocks, we propose a new strategy to eliminate or reduce the packet reordering seen by TCP. The new strategy, Sorting Reordered Packets with Interrupt Coalescing (SRPIC), works in the network device driver of the receiver. It makes use of the interrupt coalescing mechanism to sort the reordered packets belonging to each TCP stream in the interrupt coalesced blocks before delivering them. Each block is then internally free of reordering. We are focusing on the TCP packet reordering in this paper; however SRPIC could also be applied to other services.

Clearly, for our proposed strategy to work, two questions need to be answered first:

(1) Does sorting the reordered packets within blocks eliminate or reduce packet reordering?

(2) Is the sorting of reordered packets in the network device driver more efficient than processing them in the TCP layer?

In the following sections, we first give an example[†] to illustrate the effectiveness of sorting the reordered packets in blocks to eliminate or reduce packet reordering. The second question will be answered after the SRPIC algorithm is presented.

Consider that 20 packets are sent from a TCP sender to a TCP receiver, and those packets arrive in the receiver in the order as shown in Figure 3.A. According to the packet reordering metrics in [12], there are 6 packets reordered in Figure 3.A, yielding a packet reordering rate of 30%, and maximum packet reordering extent of 3. Therefore, if the network device driver delivers the received packet upward in the same sequence as they were received from the network, the protocol stack will deal with those 6 instances of packet reordering in the TCP layer. In the Figure, the packet reordering instances are highlighted in red.

---

[†] Because the packet reordering occurs randomly, it is difficult to perform the purely mathematical analysis.

Now assume that the original packet sequence in Figure 3.A can be sorted in blocks. Figure 3.B and 3.C give the resulting packet sequence with a sorting block size of 5 and 10 respectively. We summarize the packet reordering metrics for the resulting packet sequences in Table 1, and compare them with the original one. Before continuing, we give two definitions:

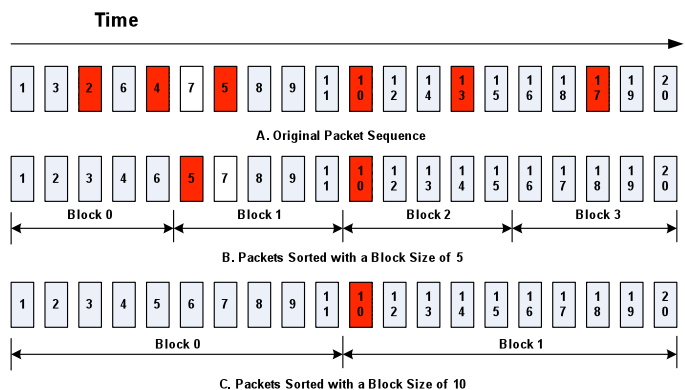

**Figure 3 Sorting Packet Reordering in Blocks**

- *Intra-block packet reordering*: packet reordering that occurs within a sorting block.
- *Inter-block packet reordering*: packet reordering that occurs across sorting blocks.

In Figure 3.B, intra-block packet reordering includes packet 2, 4, 13, and 17; inter-block packet reordering includes packet 5 and 10. In Figure 3.C, intra-block packet reordering includes packet 2, 4, 5, 13, and 17; inter-block packet reordering includes only packet 10. Figure 3 and Table 1 clearly show that sorting the reordered packet sequence in blocks can effectively eliminate intra-block packet reordering, but does not eliminate inter-block packet reordering, although it might reduce the packet reordering extent (e.g., Packet 5 in Figure 3.B). With larger sorting block sizes, reordered packets have more chance of belonging to *Intra-block packet reordering*. The conclusion is that sorting the reordered packet sequence in blocks can effectively eliminate or reduce the overall packet reordering ratio. In general, the larger the block size, the better the effect.

| Sorting Block Size | Reordering Incidents | Reordering ratio | Maximum Reordering Extent [12] |
|---|---|---|---|
| 1 (No Sorting) | 6 | 30% | 3 |
| 5 | 3 | 10% | 1 |
| 10 | 1 | 5% | 1 |

**Table 1 Packet Reordering Metrics with Different Sorting Block Sizes**

### 3.3 SRPIC Algorithm & Implementation

The interrupt coalescing mechanism handles multiple packets per interrupt as the packet rate increases. In each interrupt, the packets belonging to the same stream naturally make up a SRPIC sorting block. We have implemented the proposed SRPIC in the Linux. In order to implement SRPIC, the network device driver has a *SRPIC_manager* to administer each TCP stream, which is differentiated by the combination of *{src ip_addr, dst ip_addr, src tcp_port, dst tcp_port}*. Each *SRPIC_manager* has a structure which looks like:

```
STRUCT SRPIC_manager {
…
int     BlockSize;        /* Maximum block size to sort packet reordering */
int     PacketCnt;        /* The number of packets in the block */
int     NextExp;          /* The next expected sequence number in the receiver. The stored value in NextExp is
                             determined from a previous packet */
List    *PrevPacketList;  /* The packet list for out-of-order packets with sequence numbers less than NextExp */
List    *CurrPacketList;  /* The packet list for in-sequence packets */
List    *AfterPacketList; /* The packet list for out-of-order packets with sequence numbers larger than NextExp */
…
}
```

The *SRPIC_manager* is dynamically created or destroyed. When created or reinitialized, all the elements of *SRPIC_manager* will be set to zero. When the network device driver fetches a packet from the ring buffer, it first checks whether the packet is suitable for SRPIC. It is clear that fragmented and non-TCP packets are not appropriate. However, TCP packets (segments) are also not suitable for SRPIC if their headers include IP or TCP options (except the timestamp option), or if their TCP control bits are set (e.g., ECE, CWR, URG, RST, SYN, and FIN). These packets might need special and immediate treatment by higher layer protocols and should not be held back in the network device driver. Packets not suitable for SRPIC are delivered upward as usual. In general, the pseudo code for the SRPIC algorithm works as shown in Listing 1.

```
static int  Global_SRPIC_PacketCnt = 0;
while (not ring_buffer_is_empty ())
{
    P = Fetch_packet_from_ringbuffer ();
    if (not Packet_suitable_for_SRPIC (P))   Deliver_packet_upwards (P);
    else {
        if ((M = Find_SRPIC_manager_for_packet (P)) == NULL)
             M = Create_SRPIC_manager_for_packet (P);
        if (M→PacketCnt == 0) {
             M→NextExp = TCP_payload_1st_byte_sequence (P) +TCP_payload_len (P);
             M→PacketCnt ++;
             Add_packet_to_list_tail (P, M→CurrPacketList);
        } else {
             if (TCP_payload_1st_byte_sequence (P) < M→NextExp)
                  Add_packet_to_list_and_sort (M→PrevPacketList, P);
             if (TCP_payload_1st_byte_sequence (P) == M→NextExp) {
                  Add_packet_to_list_tail (M→CurrPacketList, P);
                  M→NextExp = TCP_payload_1st_byte_sequence (P) + TCP_payload_len (P);
             }
             if (TCP_payload_1st_byte_sequence (P) > M→NextExp)
                  Add_packet_to_list_and_sort (M→AfterPacketList, P);
             M→PacketCnt++;
             If (M→PacketCnt >= M→BlockSize) Flush_SRPIC_manager (M);
        }
        Global_SRPIC_PacketCnt++;
        if(Global_SRPIC_PacketCnt >= Ringbuffer_Size) {
             Flush_all_SRPIC_managers();
             Global_SRPIC_PacketCnt = 0;
} } }
Flush_all_SRPIC_managers ();
Global_SRPIC_PacketCnt = 0;
```

**Listing 1 Pseudo Code for SRPIC implementation**

As shown in the pseudo code, SRPIC has three packet lists: *PrevPacketList*, *CurrPacketList*, and *AfterPacketList*. The first packet in each sorting block will always go to *CurrPacketList*; then *NextExp* is updated to *TCP_payload_1st_byte_sequence (P) +TCP_payload_len (P)*. Here, *TCP_payload_1st_byte_sequence (P)* obtains the first byte sequence number of *P*'s payload; and *TCP_payload_len (P)* calculates *P*'s payload length. After that, incoming packets will be delivered to different packet lists depending on whether they are in-sequence or not. SRPIC compares the first byte sequence number of an incoming packet's payload with *NextExp;* if equal, the packet is in-sequence. Then, it is added to the *CurrPacketList* and *NextExp* is correspondingly updated. Otherwise, the packet is out-of-sequence, which will be delivered to either *PrevPacketList* or *AfterPacketList*. Packets loaded into *PrevPacketList* and *AfterPacketList* are sorted. Table 2 gives an example to illustrate the packet list operations. Assuming each packet's payload length is 1, the sorting block consists of 7 packets, and packet arrival sequence is: 2→3→1→4→6→7→5.

| *PrevPacketList: {}* <br> *CurrPacketList: {}* <br> *AfterPacketList:{}* <br> *NextExp: 0* <br><br> **Step 0: Init** | *PrevPacketList: {}* <br> *CurrPacketList: {2}* <br> *AfterPacketList:{}* <br> *NextExp: 3* <br><br> **Step 1: 2 arrives** | *PrevPacketList: {}* <br> *CurrPacketList: {2,3}* <br> *AfterPacketList:{}* <br> *NextExp: 4* <br><br> **Step 2: 3 arrives** | *PrevPacketList: {1}* <br> *CurrPacketList: {2,3}* <br> *AfterPacketList:{}* <br> *NextExp: 4* <br><br> **Step 3: 1 arrive** |
|---|---|---|---|
| *PrevPacketList: {1}* <br> *CurrPacketList: {2,3,4}* <br> *AfterPacketList:{}* <br> *NextExp: 5* <br><br> **Step 4: 4 arrives** | *PrevPacketList: {1}* <br> *CurrPacketList: {2,3,4}* <br> *AfterPacketList:{6}* <br> *NextExp: 5* <br><br> **Step 5: 6 arrives** | *PrevPacketList: {1}* <br> *CurrPacketList: {2,3,4}* <br> *AfterPacketList:{6,7}* <br> *NextExp: 5* <br><br> **Step 6: 7 arrives** | *PrevPacketList: {1}* <br> *CurrPacketList: {2,3,4,5}* <br> *AfterPacketList:{6,7}* <br> *NextExp: 6* <br><br> **Step 7: 5 arrives** |

**Table 2 SRPIC Packet Lists Operation**

Clearly, the fate of the subsequent packets in a sorting block depends significantly on the sequence number of the first packet. This design is elegant in its simplicity: SRPIC is stateless across sorting blocks and SRPIC is computable "on the fly". The purpose of having three packet lists is to reduce the sorting overheads: packets will normally arrive in sequence; the *NextExp* and *CurrPacketList* will ensure that most packets will be placed at the tail of *CurrPacketList* without being sorted. Another advantage of this implementation is that Large Receive Offload (LRO) [36] can be performed on *CurrPacketList*. We do not suggest LRO be implemented in *PrevPacketList* or *AfterPacketList*. Because these two lists keep non-in-sequence packets, there might be holes between neighboring packets in the lists, making LRO's overheads too high.

As has been discussed in previous sections, the larger the block size, the better the effect of reducing or eliminating packet reordering ratio and extent. But if all packets were delivered upward only at the end of interrupt coalescing (emptying the ring buffer), the block size might be large and the early packets in a sorting block might be delayed too long for higher layer processing, degrading performance. Let's continue the mathematical analysis in Section 3.1. Assume SRPIC is in operation and there is no constraint on the size of interrupt-coalesced blocks, the software interrupt *softnet*'s packet service rate is now $R'_{sn}$ (pps); $P(t)$ is relatively stable in cycle $j$, with average value $P^j$, and then we will have

$$T^j_{sn} = \left\lceil \frac{1 + T_{intr} * P^j}{R'_{sn} - P^j} * R'_{sn} \right\rceil / R'_{sn} \qquad (4)$$

$T^j_{sn}$ is the time that softnet spends emptying ring buffer in cycle $j$ and also the extra delay that the first packet of block $j$ incurs due to SRPIC. If $P^j$ is high, $T^j_{sn}$ could be large and the early packets in a sorting block might be delayed long. To prevent this, *BlockSize* controls the maximum block size for SRPIC_manager, and then we will have

$$\forall j > 0, \ T^j_{sn} \leq BlockSize / R'_{sn} \qquad (5)$$

*BlockSize* is configurable. The default value for *BlockSize* is 32, which is large enough to eliminate mild to moderate packet reordering. When network packet reordering is severe, it can be configured relatively large. With current computing power, $T^j_{sn}$ is usually at microsecond level, its effect on RTT could be ignored. When the number of accumulated packets in a *SRPIC_manager* reaches *BlockSize*, *Flush_SRPIC_manager ()* will deliver the packet block for higher layer processing in the sequence: *PrevPacketList, CurrPacketList, AfterPacketList,* and then the *SRPIC_manager* will be reinitialized.

SRPIC prevents high throughput connections from preempting idle connections like Telnet or SSH. SRPIC has a global variable *Global_SRPIC_PacketCnt* to counts the amount of packets for SRPIC sorting before a full flush of all *SRPIC_managers*. When *Global_SRPIC_PacketCnt* reaches *Ringbuffer_Size, Flush_all_SRPIC_managers()* will send all the packets upward even if the number of accumulated packets within a *SRPIC_manager* does not reach *BlockSize*. *Ringbuffer_Size* is the receive ring buffer size. It is a design parameter for the NIC and driver. For example, Myricom 10G NIC's is 512, Intel's 1G NIC's is 256. As such, the maximum extra delay that idle connections can experiences is $Ringbuffer\_Size / R_{sn}^{'}$. With current computing power, this is usually at most at sub-millisecond's level and can be neglected for idle connections.

At the end of interrupt coalescing, *Flush_all_SRPIC_managers()* will send all the packets upward even if the number of accumulated packets within a *SRPIC_manager* does not reach *BlockSize*. This limits delays in packet delivery.

Now let's deal with the second question raised in Section 3.2—is this efficient? Our answer is a definite yes! This is because for connection-oriented reliable data transmission, the function of sorting reordered packets is necessary and can not be avoided. If the packet reordering is eliminated or reduced in the network device driver, work in the TCP layer, which acts on each packet in the order presented, is saved. The implementation of SRPIC itself does not take much overhead: managing a *SRPIC_manager* for each TCP stream, checking if an incoming packet is suitable for SRPIC; and manipulating the three packet lists. However, the savings in the TCP layer could be significant. All the negative impact of packet reordering on TCP discussed in Section 2.2 could be saved or reduced.

Another big advantage of SRPIC is that it cooperates with any existing TCP-level packet ordering tolerant algorithms to enhance the overall TCP throughput.

We assert that sorting of reordered packets in the network device driver is much more cost-efficient than dealing with packet reordering in the TCP layer. In section 4, we will further verify the claim through experimentation. However, it needs to emphasize that SRPIC is a mechanism in the device driver that complements the TCP layer, instead of replacing it; packet reordering that can not be eliminated by SRPIC will be finally processed by TCP.

## 4. Experiment and Analysis

To verify our claims in previous sections, we run data transmission experiments upon the testing networks shown in Figure 4. The testing networks consist of a sender, a network emulator, a network switch and a receiver. In our experiments, *iperf* is sending from the sender to the receiver via the network emulator. The network emulator is a Linux system, acting as a router to forward traffic between the sender and the receiver. The network emulator has two 1Gbps interfaces, eth0 and eth1. To emulate various network conditions (e.g. delay, traffic drop, reordering etc), Netem [37] is configured on both interfaces, eth1 to emulate the forward path and eth0 to emulate the reverse path. During the experiments, the background traffic in the network is kept low. Without extra delay configured by Netem, the Round Trip Time (RTT) statistics between the sender and the receiver are: *min/avg/max/dev = 0.134/0.146/0.221/0.25 ms*. There is no packet loss or reordering in the network, except the intended packet reordering and drops configured by Netem in the experiments. The system features for the experiments are as shown in Table 3.

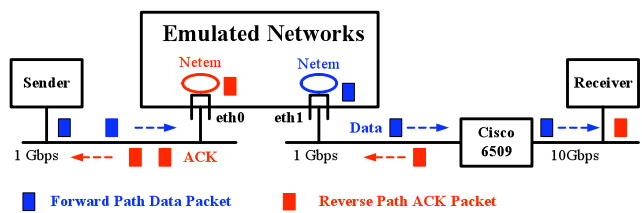

**Figure 4 Experiment Networks**

|  | Sender | Network Emulator | Receiver |
|---|---|---|---|
| **CPU** | Two Intel Xeon E5335 CPUs, 2.00 GHz, (Family 6, Model 15) | Two Intel Xeon E5335 CPUs, 2.00 GHz, (Family 6, Model 15) | Two Intel Xeon CPUs, 3.80 GHz, (Family 15, Model 4) |
| **NIC** | Intel PRO/1000 1Gbps, twisted pair | Two Intel PRO/1000 1Gbps, twisted pair | Myricom-10G PCI-Express x8, 10Gbps |
| **OS** | Linux 2.6.25, Web100 patched | Linux 2.6.24 | Linux 2.6.24 |

**Table 3 Experiment System Features**

The latest Linux network stack supports an adaptive TCP reordering threshold mechanism[‡]. Under Linux, *dupthresh* is adaptively adjusted in the sender to reduce unnecessary retransmissions and spurious congestion window reduction. It can reach as large as 127. But some network stacks, such as Windows XP and FreeBSD, still implement a static TCP reordering threshold mechanism with a default *dupthresh* value of 3. Since SRPIC is an active packet-reordering reducing mechanism on the receiver side, both adaptive and static TCP reordering threshold mechanisms on the sender side are employed in our experiments. To simulate the static threshold, we modified Linux by fixing *dupthresh* at 3. For clarity, a sender with adaptive TCP reordering threshold is termed an A-Sender, while a sender with a static reordering threshold is termed an S-Sender.

We run TCP data transmission experiments from sender to receiver. Unless otherwise specified, the sender transmits one TCP stream to the receiver for 50 seconds. In our experiments, the TCP congestion control in the sender is CUBIC [14]. We vary the network conditions in the forward and reverse paths respectively. Under the same experiment conditions, the experiment results with a SRPIC receiver are compared to those with a Non-SRPIC receiver. The metrics of interest are: (1) *Throughput (Mbps)*; (2) *PktsRetrans*, number of segments transmitted containing at least some retransmitted data; (3) *DupAcksIn*, number of duplicate ACKs received; (4) *SackBlocksRcvd*, number of SACKs blocks Received. To obtain these experiment metrics, the sender is patched with Web100 software [38]. For better comparison, *PktsRetrans, DupAcksIn,* and *SackBlocksRcvd* are normalized with *throughput (Mbps)* as: *PktsRetrans/Mbps, DupAcksIn/Mbps, and SackBlocksRcvd/Mbps.* Consistent results are obtained across repeated runs. All results presented in the paper are shown with 95% confidence interval.

---

[‡]To support such a mechanism, the reordering-tolerant algorithms implemented in Linux include Eifel algorithm, DSACK, and RFC 3517.

### 4.1 The Forward Path Experiments

### 4.1.1 Reordering experiments

In the experiments, path delays are added both in the forward and reverse paths. In the forward path, the delay follows the normal distribution; the mean and standard deviation of the path delay are $\alpha$ and $\beta \times \alpha$, respectively, where $\beta$ is the relative variation factor. A larger $\beta$ induces more variation in the path delay, hence increasing the degree of packet reordering. In the reverse path, the delay is fixed at $\alpha$. No packet drops are induced at this point. Also, SACK is turned off to reduce its influence. In the following sections, when not otherwise specified, SACK is turned on if there are packet drops introduced in the experiments. In the experiments, $\alpha$ is 2.5 *ms* and $\beta$ is varied. The sender transmits multiple parallel TCP streams (1, 5, and 10 respectively) to the receiver for 50 seconds. The results are as shown Table 4.

From Table 4, it can be seen that: (1) SRPIC can effectively increase the TCP throughput under different degrees of packet reordering (except the cases that throughput saturates the 1Gbps link), for both S-Sender and A-Sender. For example, at $\beta$ = 0.2% with 1 TCP stream, SRPIC surprisingly increases the TCP throughput more than 100% for S-Sender. (2) SRPIC significantly reduces the packet retransmission for S-Sender. (3) SRPIC effectively reduces the packet reordering in the receiver; the duplicate ACKs to the sender are significantly reduced.

| | | $\beta$ | Throughput (Mbps) | | PktsRetrans/Mbps | | DupAcksIn/Mbps | |
|---|---|---|---|---|---|---|---|---|
| | | | N-SRPIC | SRPIC | N-SRPIC | SRPIC | N-SRPIC | SRPIC |
| S-Sender | 1 Stream | 10% | 59.7 ± 0.1 | 62.5 ± 0.1 | 1094.2 ± 5.1 | 1061.5 ± 8.83 | 2959.9 ± 0.9 | 2882.1 ± 2.9 |
| | | 2% | 132.0 ± 2.2 | 176.7 ± 1.7 | 1014.2 ± 0.6 | 771.6 ± 0.6 | 2252.1 ± 2.7 | 1742.2 ± 4.2 |
| | | 1% | 145.3 ± 5.1 | 240.3 ± 1.3 | 985.1 ± 3.9 | 553.0 ± 1.6 | 2148.9 ± 16.8 | 1242.3 ± 2.3 |
| | | 0.2% | 225.0 ± 14.4 | 472.0 ± 4.1 | 794.6 ± 5.8 | 311.6 ± 4.1 | 1660.1 ± 14.6 | 694.1 ± 8.0 |
| | 5 Streams | 10% | 279.7 ± 0.7 | 300.0 ± 1.1 | 1068.9 ± 1.8 | 1028.9 ± 4.5 | 2917.8 ± 1.2 | 2811.2 ± 3.5 |
| | | 2% | 466.3 ± 2.8 | 562.0 ± 3.0 | 999.4 ± 0.5 | 848.6 ± 7.9 | 2371.5 ± 3.7 | 2035.0 ± 5.5 |
| | | 1% | 518.7 ± 4.6 | 634.3 ± 1.7 | 971.7 ± 0.9 | 814.4 ± 0.9 | 2306.2 ± 3.1 | 1906.8 ± 5.6 |
| | | 0.2% | 759.6 ± 2.3 | 876.3 ± 4.6 | 782.0 ± 1.3 | 347.1 ± 27.3 | 1705.4 ± 2.2 | 813.7 ± 25.8 |
| | 10 Streams | 10% | 536.7 ± 18.3 | 585.3 ± 0.7 | 1058.9 ± 21.5 | 1036.5 ± 1.2 | 2897.5 ± 32.7 | 2818.4 ± 1.9 |
| | | 2% | 759.7 ± 0.7 | 776.7 ± 0.7 | 1041.4 ± 1.3 | 932.6 ± 4.4 | 2374.6 ± 2.7 | 1988.1 ± 6.1 |
| | | 1% | 762.7 ± 0.7 | 787.0 ± 5.9 | 1023.5 ± 4.0 | 847.8 ± 87.3 | 2253.7 ± 18.2 | 1800.7 ± 79.8 |
| | | 0.2% | 804.0 ± 3.0 | 894.7 ± 3.5 | 768.1 ± 20.7 | 248.2 ± 20.3 | 1662.8 ± 31.8 | 706.9 ± 30.5 |
| A-Sender | 1 Stream | 10% | 338.7 ± 0.7 | 440.7 ± 0.7 | 0.5 ± 0.1 | 0.4 ± 0.1 | 3344.3 ± 5.1 | 3349.0 ± 2.8 |
| | | 2% | 800.0 ± 1.1 | 944.0 ± 0.0 | 0.3 ± 0.1 | 0.4 ± 0.3 | 3031.2 ± 5.3 | 2490.9 ± 57.9 |
| | | 1% | 944 ± 0.0 | 944.0 ± 0.0 | 0.6 ± 0.4 | 0.3 ± 0.3 | 2817.9 ± 24.1 | 2268.9 ± 13.4 |
| | | 0.2% | 944.6 ± 1.3 | 944.3 ± 0.7 | 0.8 ± 0.7 | 0.5 ± 0.3 | 1601.1 ± 41.7 | 744.1 ± 6.6 |
| | 5 Streams | 10% | 943.3 ± 0.7 | 944.0 ± 0.0 | 1.3 ± 0.1 | 1.2 ± 0.4 | 2940.3 ± 2.8 | 2803.9 ± 2.5 |
| | | 2% | 945 ± 0 | 944.0 ± 3.0 | 0.7 ± 0.1 | 0.8 ± 0.4 | 1986.8 ± 11.7 | 1598.6 ± 18.6 |
| | | 1% | 944.3 ± 0.7 | 943.0 ± 3.9 | 0.6 ± 0.1 | 0.7 ± 0.3 | 1821.0 ± 42.8 | 1418.2 ± 70.9 |
| | | 0.2% | 945.3 ± 0.7 | 947.3 ± 0.7 | 0.9 ± 0.1 | 0.5 ± 0.1 | 1169.7 ± 21.3 | 535.6 ± 26.8 |
| | 10 Streams | 10% | 944.3 ± 0.7 | 945.0 ± 0.0 | 1.2 ± 0.02 | 1.2 ± 0.1 | 2322.2 ± 6.3 | 2203.6 ± 25.1 |
| | | 2% | 944.7 ± 0.7 | 945.3 ± 0.7 | 0.7 ± 0.3 | 0.6 ± 0.2 | 1611.6 ± 3.7 | 1304.9 ± 15.3 |
| | | 1% | 944.7 ± 0.7 | 945.3 ± 0.7 | 0.8 ± 0.4 | 0.5 ± 0.1 | 1575.8 ± 48.9 | 1195.8 ± 40.9 |
| | | 0.2% | 945.3 ± 0.7 | 945.3 ± 1.3 | 0.9 ± 0.1 | 0.5 ± 0.02 | 1164.6 ± 15.2 | 537.7 ± 4.4 |

Table 4 Reordering Experiments: $\alpha = 2.5ms$ and $\beta$ is varied

Although SRPIC can significantly reduce the packet reordering in the receiver, it is interesting to note that: A-Sender does few packet retransmissions for both SRPIC and Non-SRPIC cases, the effect of SRPIC in reducing packet retransmission for A-Sender is not as significant as S-Sender; The effect of SRPIC in improving the throughput for A-Sender is not as significant as S-Sender. This is due to the following facts: (1) for S-Sender, *dupthresh* is static and fixed at 3; three consecutive duplicate ACKs will lead to fast retransmission and unnecessarily reduce congestion window in the sender. Since SRPIC can significantly reduce duplicate ACKs, chances of unnecessarily invoking packet retransmissions and spurious reduction of congestion window in the sender may be significantly reduced with SRPIC. (2) A-Sender automatically detects packet reordering and adaptively adjusts its *dupthresh* ($\geq 3$). We have noticed that A-Sender's *dupthresh* can reach as large as 127 in the reordering experiments. When *dupthresh* is adjusted high, the chance of unnecessarily invoking packet retransmission and spurious reduction of congestion window in the sender is considerably reduced even with significant amounts of duplicate ACKs to the sender. Therefore, the effect of SRPIC in improving the throughput and reducing packet retransmission for A-Sender is not as significant as S-Sender. However, it should be emphasized that the adaptive TCP reordering threshold mechanism is a sender side mechanism; it can not reduce the packet reordering of the forward path seen by receiver TCP as SRPIC does. The experiments supports the claim that SRPIC can cooperate with existing packet ordering tolerant algorithms to enhance the overall TCP performance.

From Table 4, it is apparent that the effect of SRPIC is more significant when $\beta$ is smaller. With a smaller $\beta$, more reordered packets belong to intra-block packet reordering, and SRPIC effectively eliminates intra-block packet reordering. When the degree of the packet reordering is high, more packets are inter-block reordered; SRPIC can not reduce inter-block packet reordering. As shown in Table 4, the difference of DupAcksIn/Mbps between Non-SRPIC and SRPIC when $\beta = 10\%$ is not as significant as that with $\beta = 0.2\%$. However, it still can be seen that SRPIC can increase the TCP throughput at $\beta = 10\%$. This is because SRPIC eliminates the intra-block packet reordering; even small, it still increases the chances of opening up sender's congestion window *cwnd* more than Non-SRPIC would.

The actual sorting block sizes of SRPIC are decided by incoming data rate and *BlockSize*. *BlockSize* controls the maximum sorting block size for SRPIC. However, if incoming data rate is low, the size of an interrupt coalesced block might not reach *BlockSize;* as such, the SRPIC sorting block size is purely decided by incoming data rate. It is clear that low incoming data rate will lead to small SRPIC sorting blocks. Table 4 clearly shows that SRPIC is less effective when the throughput rates are low.

With multiple concurrent TCP streams, the effective bandwidth available to each stream is actually reduced. However, the multiple-stream experiments demonstrate the effectiveness of SRPIC in reducing the packet reordering of the forward path seen by receiver TCP. This is because that TCP traffic is bursty and the instantaneous data rate for each stream could still be high. As such, each stream can still form effective SRPIC sorting blocks.

Table 4 shows the effectiveness of SRPIC in reducing the packet reordering of the forward path seen by receiver TCP, and validate the benefits of reducing packet reordering for TCP claimed previously: (1) Save costly processing in the TCP layer of both the sender and the receiver. The generation and transmission of DupACKs in the receiver are reduced, as is the processing time for DupACKs in the sender. The chance of staying in the fast path increases. The sender's system efficiency is enhanced. But we still can not claim that the receiver's efficiency is enhanced: the experiments cannot demonstrate that the savings in TCP compensates SRPIC overheads in the receiver. We will prove the enhancement of receiver's efficiency in Section 4.1.4. (2) Achieve higher TCP throughput. This is because the chance of unnecessarily reducing congestion window in the sender significantly decreases with SRPIC. (3) Maintain TCP self-clocking, and avoid injecting bursty traffic into the network. (4) Reduce or eliminate unnecessary traffic retransmission and DupACKs/SACKs traffic in the network, enhancing the overall network efficiency.

#### 4.1.2 Packet drop experiments

Path delays are added in both the forward and reverse paths, with the delay fixed at $\alpha = 2.5\,ms$ in both directions. Packet drop are added in the forward path, uniformly distributed with a variable rate of $\delta$. The purpose of packet drop experiments is to evaluate whether they have negative impact on SRPIC performance. Since packet drops will cause gaps in the traffic stream, SRPIC would hold and sort the out-of-sequence packets which are actually not misordered. The experiments are run with one TCP stream; no packet reordering is induced except the retransmission reordering caused by packet drops. SACK is enabled in the experiments.

The results are as shown in Table 5. It can be seen that packet drops will not cause negative impact on SRPIC performance. The SRPIC throughputs are almost the same with those of Non-SRPIC at different $\delta$ levels for both S-Sender and A-Sender. Similar results are observed for other metrics like PktsRetrans, DupAcksIn, and SackBlocksRcvd. We believe that this is because that the *BlockSize* and *Flush_all_SRPIC_managers()* guarantee the packets held for sorting get delivered upward in timely manner. Compared to RTT (milliseconds), the SRPIC processing delays (microseconds) is negligible and should not cause any negative effect on the performance.

In real networks, packet drops, such as congestion-induced packet loss, are usually bursty. We also run experiments with bursty packet drops. Similar conclusions as above can be drawn. Due to space limitations, those results are not included here. In the following sections, packet drops are also uniformly distributed. Experiments with multiple TCP streams draw similar conclusions.

|  | $\delta$ | Throughput (Mbps) | | PktsRetrans/Mbps | | DupAcksIn/Mbps | | SackBlocksRcvd/Mbps | |
| --- | --- | --- | --- | --- | --- | --- | --- | --- | --- |
|  |  | N-SRPIC | SRPIC | N-SRPIC | SRPIC | N-SRPIC | SRPIC | N-SRPIC | SRPIC |
| S-Sender | 0.1% | 35.7 ± 1.0 | 36.1 ± 3.3 | 4.3 ± 0.5 | 4.3 ± 0.5 | 31.8 ± 4.7 | 31.3 ± 3.1 | 32.9 ± 4.9 | 32.7 ± 2.8 |
| S-Sender | 0.01% | 139.3 ± 21.9 | 140.7 ± 17.1 | 0.4 ± 0.04 | 0.4 ± 0.1 | 12.6 ± 2.6 | 11.9 ± 2.5 | 13.1 ± 3.2 | 12.6 ± 3.5 |
| S-Sender | 0.001% | 561.6 ± 22.4 | 563.3 ± 27.2 | 0.04 ± 0.002 | 0.0 ± 0.0 | 4.3 ± 0.1 | 4.8 ± 0.7 | 4.3 ± 0.1 | 4.8 ± 0.8 |
| A-Sender | 0.1% | 37.6 ± 1.3 | 36.9 ± 0.3 | 4.4 ± 0.3 | 4.1 ± 0.3 | 34.9 ± 1.9 | 29.5 ± 2.9 | 36.5 ± 2.1 | 30.7 ± 3.3 |
| A-Sender | 0.01% | 130.0 ± 18.1 | 129.0 ± 5.9 | 0.5 ± 0.1 | 0.5 ± 0.02 | 10.7 ± 2.0 | 11.4 ± 1.6 | 10.9 ± 2.3 | 11.9 ± 2.3 |
| A-Sender | 0.001% | 615.3 ± 23.9 | 620.3 ± 17.1 | 0.03 ± 0.001 | 0.04 ± 0.004 | 3.8 ± 0.7 | 4.3 ± 0.5 | 3.9 ± 0.7 | 4.3 ± 0.5 |

**Table 5 Packet Drop Experiments, $\alpha = 2.5\,ms$, $\delta$ is varied, with 1 TCP stream**

### 4.1.3 Packet reordering & drop experiments

Both reordering and drops are added in the forward path. The forward path delay follows the normal distribution with mean and standard deviation of $\alpha = 2.5\,ms$ and $\beta \times \alpha = 0.4\% \times 2.5 = 0.01\,ms$, respectively. Packet drop is uniformly distributed with a ratio of $\delta$, where $\delta$ is varied. In the reverse path, the delay is fixed at $\alpha = 2.5\,ms$, with no drops; SACK is enabled. The sender transmits multiple parallel TCP streams (1, 5, and 10 respectively) to the receiver for 50 seconds. The results are as shown in Table 6. Conclusions similar to Section 4.1.1 can be drawn.

|    | $\delta$ | Throughput (Mbps) | | PktsRetrans/Mbps | | DupAcksIn/Mbps | | SackBlocksRcvd/Mbps | |
|----|----------|-------------------|----|------------------|----|-----------------|----|---------------------|----|
|    |          | N-SRPIC | SRPIC | N-SRPIC | SRPIC | N-SRPIC | SRPIC | N-SRPIC | SRPIC |
| 1  | 0.01%    | 67.3 ± 1.6 | 78.6 ± 1.4 | 125.8 ± 2.2 | 40.8 ± 1.3 | 1041.1 ± 9.0 | 478.9 ± 3.8 | 1125.8 ± 9.3 | 566.5 ± 5.7 |
|    | 0.001%   | 89.6 ± 1.4 | 103.0 ± 2.3 | 199.9 ± 10.0 | 50.4 ± 8.0 | 1135.7 ± 19.7 | 475.7 ± 10.9 | 1241.4 ± 25.3 | 565.1 ± 13.2 |
| 5  | 0.01%    | 174.3 ± 0.7 | 278.3 ± 0.7 | 157.6 ± 1.2 | 55.3 ± 1.4 | 1185.6 ± 4.8 | 467.8 ± 2.7 | 1326.2 ± 5.7 | 578.1 ± 4.0 |
|    | 0.001%   | 199.7 ± 2.8 | 336.7 ± 1.7 | 212.4 ± 8.1 | 67.6 ± 2.1 | 1266.3 ± 12.2 | 462.4 ± 4.1 | 1430.7 ± 16.4 | 577.1 ± 7.4 |
| 10 | 0.01%    | 309.0 ± 0 | 461.3 ± 2.6 | 165.5 ± 0.7 | 81.9 ± 0.9 | 1217.9 ± 1.7 | 490.8 ± 1.7 | 1377.3 ± 2.2 | 631.0 ± 2.7 |
|    | 0.001%   | 367.0 ± 1.1 | 510.3 ± 1.7 | 196.5 ± 1.8 | 114.8 ± 1.2 | 1245.9 ± 3.6 | 553.2 ± 2.8 | 1419.1 ± 5.5 | 726.5 ± 4.7 |

A. S-Sender

|    | $\delta$ | Throughput (Mbps) | | PktsRetrans/Mbps | | DupAcksIn/Mbps | | SackBlocksRcvd/Mbps | |
|----|----------|-------------------|----|------------------|----|-----------------|----|---------------------|----|
|    |          | N-SRPIC | SRPIC | N-SRPIC | SRPIC | N-SRPIC | SRPIC | N-SRPIC | SRPIC |
| 1  | 0.01%    | 207.7 ± 3.3 | 217.7 ± 6.2 | 0.5 ± 0.08 | 0.6 ± 0.1 | 1207.9 ± 8.0 | 493.8 ± 2.7 | 1445.9 ± 16.9 | 643.9 ± 7.4 |
|    | 0.001%   | 556.0 ± 8.1 | 624.7 ± 5.2 | 0.07 ± 0.01 | 0.07 ± 0.01 | 1915.3 ± 15.7 | 1134.8 ± 6.9 | 4069.4 ± 67.3 | 2777.4 ± 32.1 |
| 5  | 0.01%    | 911.0 ± 2.3 | 929.9 ± 3.0 | 0.48 ± 0.001 | 0.5 ± 0.01 | 1439.2 ± 0.9 | 907.4 ± 1.7 | 1983.6 ± 5.2 | 1263.7 ± 5.7 |
|    | 0.001%   | 942.0 ± 0.0 | 942.0 ± 0.0 | 0.09 ± 0.01 | 0.07 ± 0.02 | 1439.6 ± 15.8 | 880.4 ± 6.8 | 2008.8 ± 37.3 | 1232.1 ± 12.4 |
| 10 | 0.01%    | 941.3 ± 0.7 | 941.7 ± 0.7 | 0.5 ± 0.02 | 0.49 ± 0.01 | 1320.9 ± 10.6 | 840.0 ± 8.2 | 1710.7 ± 13.3 | 1115.0 ± 10.2 |
|    | 0.001%   | 942.0 ± 0.0 | 942.0 ± 0.0 | 0.08 ± 0.02 | 0.08 ± 0.01 | 1303.1 ± 2.5 | 817.4 ± 8.2 | 1675.4 ± 1.3 | 1092.5 ± 12.2 |

B. A-Sender

Table 6 Packet Reordering & Drop Experiments, $\alpha = 2.5\,ms$, $\beta = 0.4\%$ and $\delta$ is varied

### 4.1.4 CPU Comparison

Previous experiments have shown that SRPIC can effectively reduce packet reordering in the receiver, successfully increase the TCP throughput, and significantly reduce the *DupAcksIn* and *SackBlocksRcvd* to the sender. It is self-explanatory that the reduction of *DupAcksIn* and *SackBlocksRcvd* to the sender will save the costly *DupAcksIn* and *SackBlocksRcvd* processing in the sender, leading to higher system efficiency. We need to verify that SRPIC is also cost-efficient in the receiver. The savings in TCP need to compensate for the SRPIC overheads.

We run *oprofile* [39] to profile the system performance in both the sender and receiver. The metrics of interest are: (1) CPU_CLK_UNHALTED [40], the number of CPU clocks when not halted; (2) INST_RETIRED [40], the number of instruction retired. These two metrics evaluate the load on both systems. Since our testing systems are multi-core based, we pin the network interrupts and iperf to one specific CPU on each system, in order to make the profiling more accurate.

The CPU comparison experiments are run as a series of tests: (1) No packet reordering experiments, with a delay of 2.5 *ms* is added to both the forward and reverse paths. The purpose of this test is to evaluate whether SRPIC will degrade the system performance when there is no packet reordering. (2) Packet reordering experiments, with the experiment configuration identical to Table 4, where $\beta = 0.2\%$. These experiments are run with A-Sender. We do not apply packet drops or use S-Sender in the experiments: packet drops cause big throughput variation; with S-Sender, the throughput difference between SRPIC and Non-SRPIC cases is very large. The throughput rates in these experiments all saturate the 1Gbps link (around 940 Mbps). The experiments are designed to have the same throughput rates for the sake of better CPU comparison. In all these experiments, the sender transmits multiple parallel TCP streams (1, 5, and 10 respectively) to the receiver for 50 seconds. The results are as shown in Figure 5. In the figure, $n$R+ and $n$R- represent $n$ streams with or without reordering, respectively.

From Figure 5, it can be seen that without packet reordering, the loads on the sender and receiver are almost the same for both SRPIC and Non-SRPIC cases, with different number of parallel TCP streams. However, when there are reordered packets in the forward path, experiments show that SRPIC can enhance system efficiency for both the sender and receiver. With SRPIC, CPU_CLK_UNHALTED and INST_RETIRED on the sender and receiver are lower. This is because SRPIC can effectively reduce the packet reordering in the receiver; all the negative impacts discussed in Section 2.2 can be reduced. The experiments further support the claim that the savings in TCP compensate for the SRPIC overheads in the receiver. The conclusion is that SRPIC enhances the receiver's system efficiency.

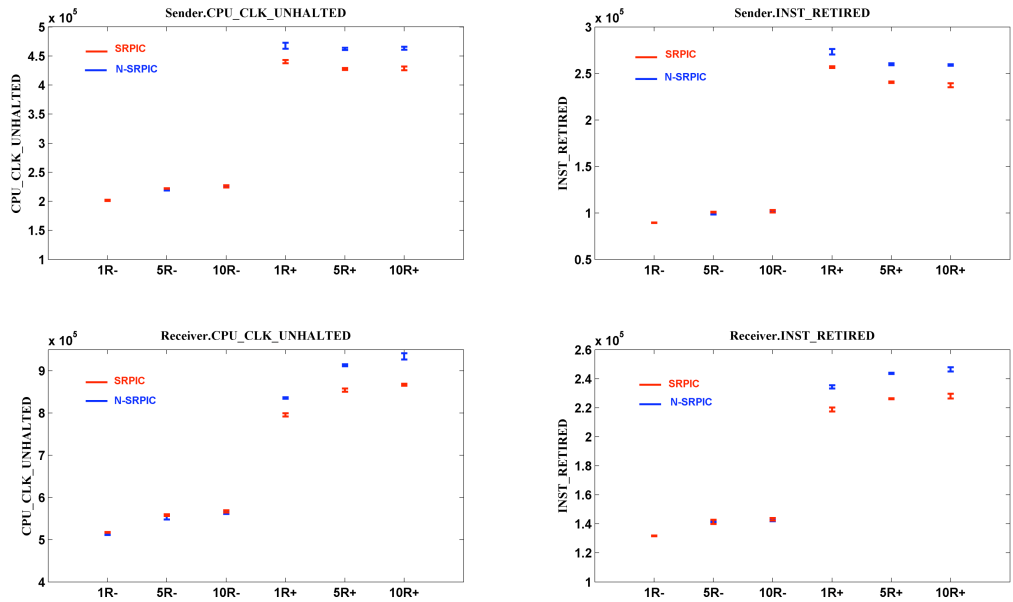

**Figure 5 CPU Comparisons, Number of Events between Samples: 100000**

### 4.2 The Reverse Path Experiments

ACKs/SACKs go back to the sender in the reverse path. Due to TCP's cumulative acknowledgements, the reverse-path reordering and packet drops can have the following impacts [4]: loss of self-clocking, and injection of bursty traffic into the network. We repeat all experiments above on the reverse path. Packet reordering and drops are added in the reverse path, with no packet reordering and drops in the forward path. The throughput rates of all these experiments saturate the 1Gbps link (around 940 Mbps). In these experiments, the metrics used to evaluate SRPIC in Section 4.1 are very close to each other between the SRPIC and Non-SRPIC cases. The results are as expected: SRPIC is the receiver side algorithm, it does not deal with the reverse path packet reordering and drops; Also, since there is no reordering or drops in the forward path, SRPIC will not take effect in the receiver. Due to space limitations, the reverse path experiment results are not included here.

**4.3 The Forward and Reverse Paths Experiments**

How will SRPIC perform when packet reordering and drops occur in both the forward and reverse paths? To answer the question, we repeat the experiments in Section 4.1.3, adding packet reordering and drops in both the forward and reverse paths. For both directions, the path delay follows the normal distribution with mean and standard deviation of $\alpha = 2.5\ ms$ and $\beta \times \alpha = 0.4\% \times 2.5 = 0.01\ ms$, respectively. Packet drop is uniformly distributed with a ratio of $\delta$. SACK is enabled.

The results are as shown in Table 7. It can be seen that SRPIC significantly reduces the packet retransmission for the S-Sender. SRPIC also effectively reduces the packet reordering in the receiver, with duplicate ACKs and SACK blocks to the sender significantly reduced as well. It is the packet reordering and drops in the forward path that leads to duplicate ACKs and SACK blocks back to the sender. The significant reduction of duplicate ACKs and SACK blocks to the sender again verify the effectiveness of SRPIC in reducing the forward path packet reordering seen by receiver TCP.

It is interesting to note that the throughput rate in Table 7 is a little bit different than in Section 4.1.3. With $\delta = 0.001\%$, SRPIC throughput rates are higher than those of Non-SRPIC, for both S-Sender and A-Sender. This is positive and as expected. However, with $\delta = 0.01\%$, the SRPIC throughput rate is slightly less than the Non-SRPIC case for A-Sender, although significant reduction of *DupAcksIn* and *SackBlocksRcvd* is still observed. We believe this phenomenon is caused by the ACK reordering in the reverse path, which somewhat negates the SRPIC's effort in reducing the forward path packet reordering. The ACK reordering should make the number of "effective" ACKs back to the sender smaller because some of them get discarded as old ones as a newer cumulative ACK often arrives a bit "ahead" of its time making rest smaller sequenced ACKs very close to a no-op. The unfortunate result is that the sender's congestion window will grow far too slowly [4]. These experiments demonstrate that combined with the forward path packet reordering, ACK reordering in the reverse path might lead to unpredictable TCP throughput, even through SRPIC can effectively reduce packet reordering in the receiver. Experiments with multiple TCP streams draw similar conclusions.

|  | $\delta$ | Throughput (Mbps) | | PktsRetrans/Mbps | | DupAcksIn/Mbps | | SackBlocksRcvd/Mbps | |
|---|---|---|---|---|---|---|---|---|---|
|  |  | N-SRPIC | SRPIC | N-SRPIC | SRPIC | N-SRPIC | SRPIC | N-SRPIC | SRPIC |
| S-Sender | 0.01% | 60.7 ± 1.4 | 64.1 ± 1.0 | 75.3 ± 0.9 | 31.3 ± 1.3 | 728.7 ± 2.0 | 445.7 ± 1.4 | 1128.3 ± 5.0 | 550.1 ± 3.5 |
| S-Sender | 0.001% | 84.4 ± 1.1 | 87.5 ± 0.7 | 103.3 ± 6.2 | 40.4 ± 2.5 | 747.6 ± 12.6 | 446.2 ± 2.4 | 1165.6 ± 18.8 | 554.2 ± 4.7 |
|  | $\delta$ | Throughput (Mbps) | | PktsRetrans/Mbps | | DupAcksIn/Mbps | | DupAcksIn/Mbps | |
|  |  | N-SRPIC | SRPIC | N-SRPIC | SRPIC | N-SRPIC | SRPIC | N-SRPIC | SRPIC |
| A-Sender | 0.01% | 183.0 ± 0.0 | 179.7 ± 3.5 | 0.4 ± 0.01 | 0.4 ± 0.02 | 752.6 ± 2.5 | 449.9 ± 2.6 | 1309.6 ± 6.3 | 575.3 ± 6.2 |
| A-Sender | 0.001% | 306.3 ± 3.6 | 356.7 ± 3.9 | 0.04 ± 0.003 | 0.05 ± 0.001 | 719.6 ± 4.3 | 352.3 ± 3.4 | 1775.9 ± 15.3 | 729.9 ± 13.8 |

**Table 7 the Forward and Reverse Paths Experiments**

### 4.4 Delay Experiments

Without doubt, SRPIC adds extra delays for TCP traffic. The TCP traffics of idle connections like Telnet or SSH are delayed the most. This is because idle connections' throughput is often low and their sorting block usually can not reach *BlockSize*. As such, the packets are delivered upward either at the end of interrupt coalescing, or when *Global_SRPIC_PacketCnt* reaches *Ringbuffer_Size*. It has been analyzed in Section 3.3 that the delay caused by SRPIC is negligible. This claim is further verified by the following experiments.

In the experiments, the sender transmits 10 parallel TCP streams to the receiver at full speeds; no packet reordering or drops are configured. At the same time, we login the receiver from a third system with SSH and type "ls –al" for a thousand time to simulate idle connection traffic. The Pinged RTT statistics between the third system and the receiver are: min/avg/max/mdev = 5.207/5.212/5.232/0.093 *ms*. The third system runs Linux 2.6.24; its

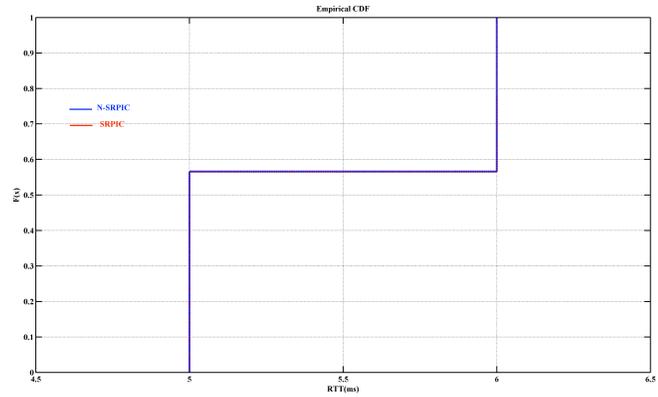

A. The third system TCP-layer calculated RTT Statistics

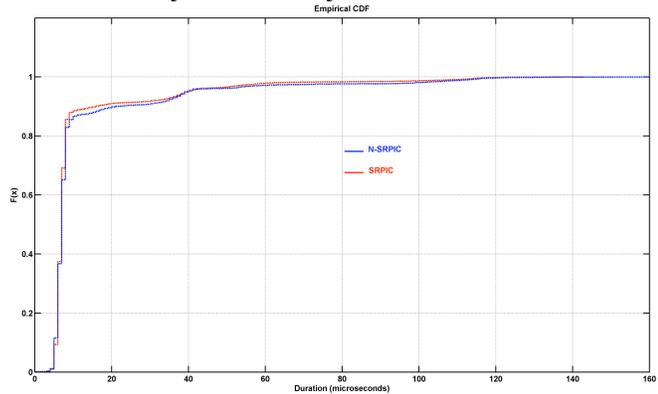

B. Receiver Softnet Emptying Ring Buffer Duration Statistics

Figure 6 Delay Experiments

system clock resolution is 1 ms. It has been instrumented to record each TCP-layer calculated RTT. The cumulative distribution functions (CDF) for the recorded TCP-layer calculated RTT with or without a SRPIC receiver are compared. Also, the receiver has been instrumented to record the duration that softnet spends emptying the ring buffer in each cycle.

The results are as shown in Figure 6. Figure 6.A shows the CDF for the third system TCP-layer calculated RTT and Figure 6.B shows the CDF for the duration that receiver softnet spends emptying the ring buffer for each interrupt coalesced cycle. Figure 6.A clearly shows that the CDF with SRPIC almost totally overlaps that of Non-SRPIC. For both cases, the TCP-layer calculated RTT is either 5ms or 6 ms. This is because TCP calculates segments' RTT with integral jiffy units. Jiffy represents system clock granularity; for Linux and other OSes, the finest system clock granularity only reaches 1 ms level. TCP-layer can not tell the RTT changes less than 1 ms. As it has been analyzed in Section 3.3, the extra delay caused by SRPIC is at most at sub-milliseconds' level. This is further verified by the experiment results of Figure 6.B. For the duration that receiver softnet spends emptying the ring buffer for each interrupt coalesced cycle, the CDF with SRPIC almost totally overlaps that of Non-SRPIC; mostly, the duration is less than 20 microseconds. It verifies our claim that the delay caused by SRPIC is negligible.

## 5. Discussion and Conclusions

Originally, we had planned to implement a SRPIC-like mechanism between the IP and TCP layers. The final implementation of SRPIC was inspired by the LRO implementation in [36]. When combined with LRO, the overheads of SRPIC are at least cut in half.

Implementing a mechanism similar to SRPIC between the IP and TCP layers has the advantages of saving the overheads in maintaining the *SRPIC_manager* for each TCP stream, where SRPIC could be performed within each TCP socket. However, there is no natural grouping of packets into blocks at that layer. A timing or other mechanism would be needed to hold packets for sorting and to flush sorted blocks at appropriate times. Still, since TCP traffic is bursty, a mechanism similar to SRPIC in between the IP and TCP layers might still be worth trying. We leave it for further study.

Due to system clock resolution issues, it is difficult to emulate networks with bandwidth beyond 1Gbps. All our experiments are run upon Gigabit networks. However, SRPIC should be more effective in higher bandwidth networks.

In this paper, we have proposed a new strategy: Sorting Reordered Packets with Interrupt Coalescing (SRPIC) to eliminate or reduce the packet reordering in the receiver. Experiments have demonstrated the effectiveness of SRPIC against forward-path reordering. The significant benefits of our proposed SRPIC include reducing processing in the TCP layer in both the sender and the receiver, higher achieved TCP throughput; maintenance of TCP self-clocking while avoiding injection of bursty traffic into the network, reduction or elimination of unnecessary retransmissions and duplicate ACKs or SACKs, and coexistence with other packet reordering-tolerant algorithms.